\documentclass[11pt]{JHEP3}
\usepackage{epsfig}
\usepackage{amsmath,amsfonts,amscd,amssymb,amsthm}

\title{Self-gravitating radiation in $AdS_d$}
\author{Vladislav Vaganov \\ D.A.M.T.P., Centre for Mathematical
Sciences, University of Cambridge, Wilberforce Road, Cambridge CB3
0WA, U.K. \\E-mail: \email{vv205@cam.ac.uk}}

\abstract{We study spherically symmetric equilibrium
configurations of self-gravitating massless thermal radiation in
asymptotically anti-de Sitter space. In $d=4$, it was shown by
Page and Phillips that there is a maximum red-shifted temperature,
maximum mass and maximum entropy. For higher central densities,
the temperature, mass and entropy undergo an infinite series of
damped oscillations, corresponding to unstable configurations.  We
extend this work to all dimensions $d\geq 3$. We find that in
$4\leq d\leq 10$, the behaviour is similar to the $d=4$ case. In
$d\geq 11$, the temperature, mass and entropy are monotonic
functions of the central density, asymptoting to their maxima as
the central density goes to infinity. In $d=3$, an exact solution
is given by a slice of the $AdS$ C-metric.}

\preprint{DAMTP-2007-66}

\begin{document}

\section{Introduction}
It is well known that the canonical ensemble is not defined in
asymptotically flat space. This is because having thermal
radiation at constant temperature at infinity is not compatible
with asymptotic flatness.  One can avoid this problem by enclosing
the system in a box, which is unphysical \cite{hawkingpage}, or by
working in asymptotically anti-de Sitter ($AdS$) space. In $AdS$
the canonical ensemble is given by a Euclidean path integral over
all matter fields and metrics which tend asymptotically
respectively to zero and to $AdS$ identified with period
$\beta=T^{-1}$ in imaginary time, where $T$ is the red-shifted
temperature \cite{hawkingpage}. In their seminal paper, Hawking
and Page \cite{hawkingpage} studied this path integral in the
semiclassical approximation, where it is dominated by classical
solutions to the Einstein equations with these boundary
conditions. Their work was later generalized to $d$ dimensions
\cite{witten}. For $T<T_0=\sqrt{(d-1)(d-3)}/2\pi l$, where $l\gg
l_{Planck}$ is the $AdS$ length, there are no black hole
solutions; thermal radiation, that is, $AdS$ periodically
identified in imaginary time, is the only admissible phase. For
$T_0<T<T_{HP}=(d-2)/2\pi l$, thermal radiation is the preferred
phase, although black holes may form and evaporate from time to
time as a result of fluctuations.  For $T>T_{HP}$, the
configuration with a large $AdS$ black hole in equilibrium with
thermal radiation has a lower free energy than thermal radiation
alone and it therefore dominates the path integral. The point
$T=T_{HP}$ marks a first order phase transition, known as the
Hawking-Page transition, between two topologically distinct
manifolds. The small $AdS$ black hole has negative specific heat
and is never a dominant phase, but it serves as a bounce mediating
the tunneling amplitude between thermal radiation and the large
$AdS$ black hole \cite{hawkingpage}.

Thus far the gravitational effect of thermal radiation has been
neglected, and therefore this picture will undergo quantum
corrections when the back-reaction is taken into account. These
correspond to the one and higher loop terms in the path integral.
An exact solution describing a black hole in equilibrium with
thermal radiation remains elusive, although this problem has
recently been conjectured to be equivalent to the quest for a
localized black hole on a brane \cite{emparan}, which is a purely
classical problem in GR. For thermal radiation alone, however, a
useful model may be obtained by treating the radiation gas as a
perfect fluid with equation of state $p=\rho/(d-1)$, where the
energy density is related to the local temperature by the
Stefan-Boltzmann law, $\rho\propto T_{loc}^d$. This amounts to
taking just the leading term in the high-temperature expansion of
the one-loop effective action \cite{dowkerkennedy} and is a good
approximation if the red-shifted temperature is much greater than
the inverse of a characteristic curvature radius of the spacetime.
For a given central density, one may solve the Einstein equations
with a $\Lambda$ term for the static spherically symmetric
equilibrium configurations of self-gravitating thermal radiation.
In $d=4$, this calculation was carried out by Page and Phillips
\cite{pagephillips}. Their key finding was that there exist
locally stable radiation configurations all the way up to a
maximum red-shifted temperature $T_{max}\gg T_{HP}$, above which
there are no solutions.\footnote{This temperature had previously
been estimated in \cite{hawkingpage}: there it is known as $T_2$.}
Thus for temperatures $T>T_{max}$, the only possible phase is a
large $AdS$ black hole in equilibrium with thermal radiation
\cite{hawkingpage}. There is also a maximum mass and maximum
entropy configuration occurring at a higher central density than
the maximum temperature configuration \cite{pagephillips}. Beyond
their peaks, the temperature, mass and entropy undergo an infinite
series of damped oscillations; configurations in this range are
unstable, although the precise onset of instability depends on the
ensemble \cite{pagephillips}. The purpose of this paper is to
generalize the analysis of \cite{pagephillips} to $d$ dimensions.

The organization is as follows. In section \ref{sec2} we write
down the governing equations for this system, which are a special
case of the Tolman-Oppenheimer-Volkoff equations. The solution
space is two-dimensional, but if we impose regular boundary
conditions at the origin we get a one-parameter set of equilibrium
configurations labeled by the central density. These solutions
must be found numerically, although useful asymptotic expressions
may be obtained when the radius is much smaller or much larger
than the $AdS$ length. We show how given a solution, one may
compute its red-shifted temperature, mass and entropy. We have
calculated these for a range of central densities in each
dimension. The results are presented in section \ref{sec3}, where
we see that in each dimension $d\geq 3$ there is a maximum
red-shifted temperature, mass and entropy. However, whereas in
$4\leq d\leq 10$ the behaviour is similar to the $d=4$ case, in
$d\geq 11$ the behaviour is different: the temperature, mass and
entropy increase monotonically with the central density, with no
oscillations. In section \ref{sec4} we briefly address the range
of validity of the perfect fluid approximation. We conclude in
section \ref{sec5} with a discussion of the broader relevance of
this work. Details of the $d=3$ case are contained in Appendix
\ref{appendix1}.  Appendix \ref{appendix2} contains an analysis of
the scale-invariant $\Lambda=0$ equations.

\section{Perfect fluid radiation in asymptotically anti-de Sitter
space}\label{sec2} We want to solve the Einstein equations with
radiation gas perfect fluid source for a static spherically
symmetric metric in $d$ spacetime dimensions, with asymptotically
$AdS$ boundary conditions. We will work in the system of units in
which $c=\hbar=k_B=1$ and $\kappa=8\pi G_N$. The following is
mostly based on \cite{pagephillips} extended to $d$ dimensions.
The metric may be written as
\begin{align}\label{metric:ansatz}
ds^2&=-e^{2\psi}V dt^2 +V^{-1} dr^2 +r^2
d\Omega_{d-2}^2\nonumber\\ V(r)&\equiv
\left(1+\frac{r^2}{l^2}-{\frac {2\kappa m(r)}{(d-2)\Sigma
{r}^{d-3}}}\right)\,,
\end{align}
where $d\Omega_{d-2}^2$ is the metric on the unit $(d-2)$-sphere,
which has volume $\Sigma$. The $AdS$ length
$l=\left(-(d-1)(d-2)/2\Lambda\right)^{1/2}$, where $\Lambda<0$ is
the cosmological constant. Written in this way,
\begin{equation}\label{eqn:mass}
\lim_{r\to\infty} m(r) =M
\end{equation}
is the total mass of the configuration, and the time coordinate
$t$ may be normalized such that
\begin{equation}\label{eqn:psinorm}
\lim_{r\to\infty} \psi(r)=0\,.
\end{equation}
The energy-momentum tensor is taken to be that of a perfect fluid
with equation of state $p=\rho/(d-1)$, where $p$ is the isotropic
pressure. When the fluid is at rest with respect to the static
frame defined by the Killing vector $\partial/\partial t$, this
takes the form
\begin{equation}\label{eqn:tmunu}
T^\mu_\nu=\frac{\rho}{d-1}\left(\delta^\mu_\nu-d\delta^\mu_0
\delta^0_\nu\right),
\end{equation}
where the rest frame energy density is related to the local Tolman
temperature $T_{loc}$ through the Stefan-Boltzmann law,
\begin{equation}\label{eqn:stefanb}
\rho=aT_{loc}^d\,.
\end{equation}
The dimension-dependent constant $a$ is proportional to the
effective number of degrees of freedom (see section \ref{sec4}).

The Einstein equations $G_{\mu\nu}+\Lambda g_{\mu\nu}=\kappa
T_{\mu\nu}$ for the metric (\ref{metric:ansatz}) and the energy
momentum tensor (\ref{eqn:tmunu}) yield the following system of
ODE's:
\begin{align}\label{eqn:tov}
\frac {dm}{dr}&=\Sigma r^{d-2}\rho\nonumber\\ \frac
{d\rho}{dr}&=-\frac{\rho d\left[(d-1)(d-3)\kappa m/\Sigma+\kappa
r^{d-1}\rho+(d-1)(d-2) r^{d-1}/l^2\right]}{(d-1)(d-2) r^{d-2}V},
\end{align}
which are a case of the Tolman-Oppenheimer-Volkoff equations in
$d$ dimensions. There is a two-parameter family of solutions, but
imposing regular boundary conditions at the origin,
\begin{equation}
\rho(0)=\rho_c\qquad m(0)=0\,,
\end{equation}
yields a one-parameter set of regular equilibrium configurations
$m(r)$, $\rho(r)$ labeled by the central density $\rho_c> 0$. In
general (\ref{eqn:tov}) must be integrated numerically. Because
these equations are singular at $r=0$, the solution must be cut
off at $r=\epsilon\ll l$, where we set $\rho(\epsilon)=\rho_c$ and
$m(\epsilon)=\Sigma\rho_c\epsilon^{d-1}/(d-1)$. The physical
answers are recovered as $\epsilon$ is taken to zero. In practice,
one takes a value of $\epsilon$ small enough for the required
level of accuracy.\footnote{It is possible to regularize the
Tolman-Oppenheimer-Volkoff equations and recast them as a
$3$-dimensional regular dynamical system on a compact state space
\cite{relstellstruc}. This has a number of advantages, but is not
essential in our analysis, see, however, Appendix
\ref{appendix2}.}

Once $m(r)$ and $\rho(r)$ (and hence the metric function $V$) are
determined, we turn to the contracted Bianchi identity
(equivalently, conservation of $T_{\mu\nu}$), which reads
\begin{equation}\label{eqn:support}
\frac{d\rho}{dr}+\rho d
\left(\frac{d\psi}{dr}+\frac{1}{2}V^{-1}\frac{dV}{dr}\right)=0\,.
\end{equation}
The red-shifted temperature is defined as
\begin{equation}\label{eqn:redshifted}
T\equiv T_{loc}|g_{tt}|^{1/2}=T_{loc}e^{\psi}V^{1/2}\,.
\end{equation}
Equation (\ref{eqn:support}) implies that $T$ is constant, as it
must be for an equilibrium configuration. It follows from
condition (\ref{eqn:psinorm}) that for large $r$ the solutions
behave as $\rho\sim V^{-d/2}\sim r^{-d}$. It also follows that
given a solution $m(r)$, $\rho(r)$ the red-shifted temperature may
be computed by taking the limit
\begin{equation}\label{eqn:temp}
T=\lim_{r\to\infty} a^{-1/d}\rho^{1/d} V^{1/2}\,.
\end{equation}
Equation (\ref{eqn:redshifted}) may then be used to determine
$\psi(r)$. Note that for the regular solutions we are considering,
equations (\ref{eqn:tov}) and (\ref{eqn:support}) imply $dm/dr>
0$, $d\rho/dr< 0$ and $d\psi/dr> 0$ for all $r>0$.  Pure $AdS$ in
global coordinates corresponds to $\rho_c=0$ and has
$m=\rho=\psi=0$ for all $r\geq 0$.

One may compute the entropy $S$ of an equilibrium configuration by
integrating the local entropy density,
\begin{equation}
s=\frac{d}{d-1}aT_{loc}^{d-1}=\frac{d}{d-1}a^{1/d}\rho^{(d-1)/d}
\end{equation}
over the proper spatial volume:
\begin{equation}\label{eqn:entropy}
S=\frac{d}{d-1}a^{1/d}\Sigma \int_0^{\infty}\rho^{(d-1)/d}
V^{-1/2} r^{d-2}\; dr\,.
\end{equation}
If the mass (\ref{eqn:mass}) and temperature (\ref{eqn:temp}) are
known for a range of central densities, the entropy may also be
obtained by integrating the first law of thermodynamics, $dM=TdS$,
starting from the $\rho_c=0$ configuration which has $M=T=S=0$.

We now go on to consider certain limits of the solutions. Observe
that equations (\ref{eqn:tov}) are invariant \cite{collins} under
the scaling
\begin{equation}\label{eqn:scale}
r \rightarrow C r,\quad \rho\rightarrow C^{-2}\rho,\quad
m\rightarrow C^{d-3}m,\quad l\rightarrow C l\,,
\end{equation}
where $C\in\mathbb{R^+}$. When $\Lambda=0$ ($l=\infty$) this is an
exact scale invariance \cite{collins2}. In this case there exists
a special self-similar solution\footnote{By this we mean the
self-similarity of the geometry as captured by the existence of a
proper homothetic vector \cite{carrcoley}. This does not
necessarily follow from the self-similarity of the matter fields
as expressed by the scaling relation (\ref{eqn:scale}) with
$\Lambda=0$. Indeed, other solutions of the $\Lambda=0$ equations
do not admit a proper homothetic vector.} given by
\begin{equation}\label{eqn:singular}
\rho=\alpha r^{-2},\quad m=\alpha\Sigma r^{d-3}/(d-3)\quad
\mbox{where}\quad \alpha=\frac{2(d-1)(d-3)\kappa^{-1}}{d^2-d+2}\,,
\end{equation}
and we take $d\geq 4$ (see Appendix \ref{appendix1} for the $d=3$
case). This solution is singular at $r=0$. It is the limit as the
central density goes to infinity of the $\Lambda=0$ solutions with
a regular origin. The latter may be computed numerically, see, for
example, \cite{sorkinwaldzhang,chavanis2} for $d=4$.  It is best
formulated (Appendix \ref{appendix2}) using variables invariant
under the scale transformation (\ref{eqn:scale}), whence equations
(\ref{eqn:tov}) may be reduced to a plane autonomous system, from
which one may derive various qualitative features and asymptotic
approximations \cite{sorkinwaldzhang,chavanis2}. For large $r$,
regular solutions will approach the self-similar solution
(\ref{eqn:singular}), which has infinite mass and is thus not
asymptotically flat. To obtain finite mass solutions in the
$\Lambda=0$ case, one must confine the radiation to an unphysical
box. If one does so, the thermodynamics (red-shifted temperature,
mass and entropy as functions of the central density) of these
configurations will be qualitatively similar to the asymptotically
$AdS$ case studied here.

For fixed $\Lambda<0$, equations (\ref{eqn:tov}) are no longer
scale invariant. However, for $r\ll l$, the terms involving
$\Lambda$ may be neglected, and the solutions approach the regular
$\Lambda=0$ solution with the same value of $\rho_c$. In the limit
as the central density goes to infinity, they approach the
singular self-similar solution (\ref{eqn:singular}) for $r\ll l$.
On the other hand, for large $r$, $\rho\sim r^{-d}$, and the first
of equations (\ref{eqn:tov}) implies that the mass of these
solutions is finite. This is due to the confining nature of the
gravitational potential in anti-de Sitter space, which acts as a
box of finite volume.

\section{Results}\label{sec3}
For each dimension, we computed a number of solutions for a range
of values of $\rho_c$, and the red-shifted temperature, mass and
entropy for each solution.  A combination of Runge-Kutta methods
were used to integrate equations (\ref{eqn:tov}) starting from
$r=\epsilon$ and terminating at $r=r_1\gg l$, and then equations
(\ref{eqn:temp}), (\ref{eqn:mass}) and (\ref{eqn:entropy}) with
infinity replaced by $r_1$ were used to evaluate the temperature,
mass and entropy. The configuration with infinite central density
was computed by imposing (\ref{eqn:singular}) as a boundary
condition at $r=\epsilon$. The errors coming from neglecting the
region $r_1<r<\infty$ may be approximated by using asymptotic
expansions \cite{pagephillips}. The accuracy of the calculation
depends on the values of $\epsilon$, $r_1$ and the error tolerance
of the routine. A range of different values were used to achieve
four-significant-figure accuracy, where greater precision was
required at higher central densities and in higher dimensions. We
work in terms of the dimensionless quantities
\begin{align}\label{eqn:dimensions}
\tilde{\rho_c}&\equiv \kappa l^2\rho_c\nonumber\\ \tilde{T}&\equiv
a^{1/d}\kappa^{1/d}l^{2/d}T\nonumber\\ \tilde{M}&\equiv \kappa
l^{3-d} M\nonumber\\ \tilde{S}&\equiv
a^{-1/d}\kappa^{(d-1)/d}l^{-(d-1)(d-2)/d}S\,.
\end{align}

\noindent We found that for all $d\geq 3$ there exists a maximum
red-shifted temperature, $T_{max}$, a maximum mass $M_{max}$ and a
maximum entropy $S_{max}$, where the maximum entropy configuration
always coincides with the maximum mass configuration. The maxima
of temperature, mass and entropy, along with the corresponding
values of $\rho_c$, are presented in Table~\ref{tab1} for $3\leq
d\leq 13$. For comparison, we also give the ratio of
$\tilde{M}_{max}$ to the mass $\tilde{M}_{HP}=(d-2)\Sigma$ of the
large $AdS$ black hole at the Hawking-Page temperature: this is
simply to illustrate the similar dimensional dependence of these
quantities (see section {\ref{sec4} for some physical
comparisons). Figs.~1-3 show how the dimensionless temperature,
mass and entropy vary with the central density for three
representative cases: $d=4$, $d=11$ and $d=3$.

\TABULAR{@{\extracolsep{\fill}}|c|cc|cccc|}{\hline & &
& & & & \\ $d$ & $\ln\tilde{\rho_c}$ & $\tilde{T}_{max}$ &
$\ln\tilde{\rho_c}$ & $\tilde{M}_{max}$ &
$\tilde{M}_{max}/\tilde{M}_{HP}$ & $\tilde{S}_{max}$\\ \hline & &
& & & &
\\ $3$ & $0$ & $2/3\approx 0.6667$ & $\ln 4\approx 1.386$ & $\pi/3\approx 1.047$ & $1/3\approx 0.3333$ & $4^{2/3}\pi\approx 7.916$\\ $4$ & $1.353$ &
$0.9479$ & $2.018$ & $11.56$ & $0.4598$ & $15.22$\\ $5$ & $2.349$
& $1.081$ & $2.650$ & $22.98$ & $0.3880$ & $25.95$ \\ $6$ &
$3.234$ & $1.145$ & $3.354$ & $37.76$ & $0.3586$ & $39.22$\\ $7$ &
$4.119$ & $1.177$ & $4.158$ & $53.39$ & $0.3444$ & $52.78$\\ $8$ &
$5.109$ & $1.190$ & $5.119$ & $66.55$ & $0.3354$ & $63.81$
\\ $9$ & $6.390$ & $1.195$ & $6.392$ & $74.50$ & $0.3278$ & $70.08$ \\ $10$ & $8.590$ & $1.195$ & $8.590$ & $76.08$ & $0.3204$ & $70.72$ \\ $11$ & $\infty$ & $1.192$
& $\infty$ & $71.81$ & $0.3129$ & $66.25$\\ $12$ & $\infty$ &
$1.188$ & $\infty$ & $63.31$ & $0.3055$ & $58.14$ \\ $13$ &
$\infty$ & $1.183$ & $\infty$ & $52.57$ & $0.2983$ & $48.15$\\
\hline}{\label{tab1}Maximum red-shifted temperature, maximum mass
and maximum entropy configurations of spherically symmetric
thermal radiation in $AdS$ in various dimensions. The quantity
$\tilde{M}_{HP}=(d-2)\Sigma$ is the mass of the large $AdS$ black
hole at the Hawking-Page temperature.}

\DOUBLEFIGURE{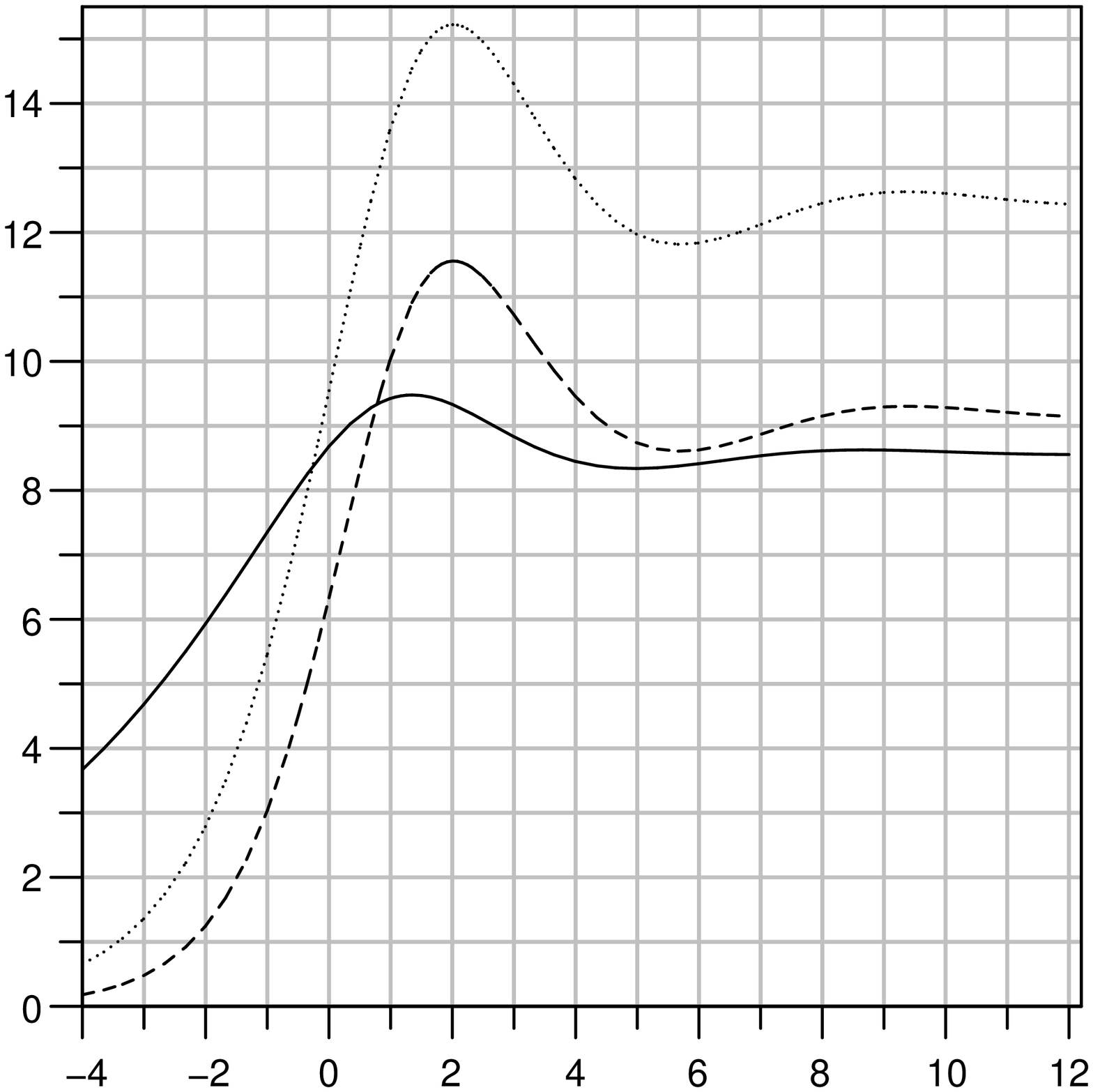,width=2.9in,height=2.9in}{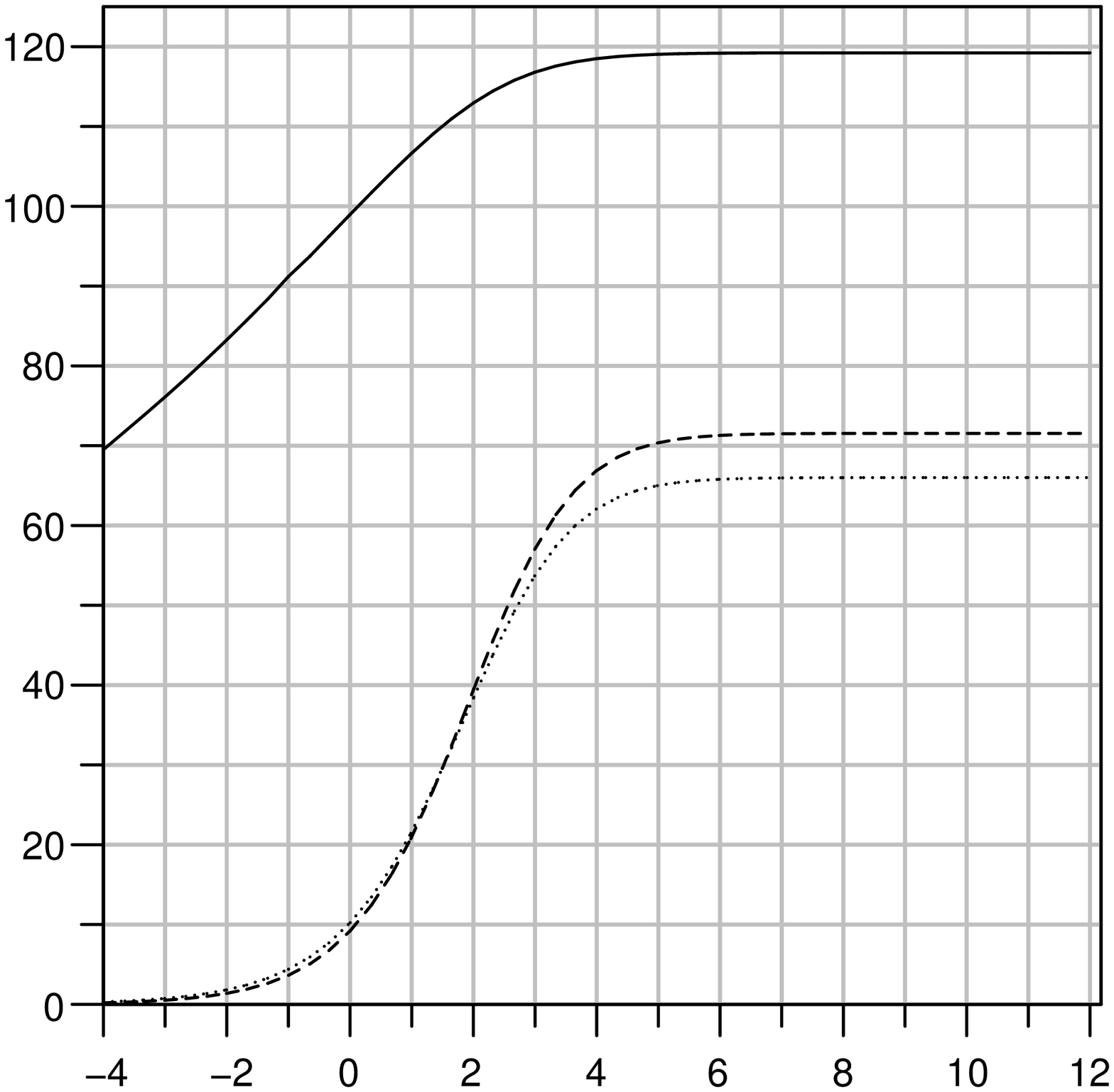,width=2.9in,height=2.9in}{\label{fig1}$d=4$:
$10\tilde{T}$ (solid line), $\tilde{M}$ (dashed line) and
$\tilde{S}$ (dotted line) vs.
$\ln\tilde{\rho_c}$}{\label{fig2}$d=11$: $10^2\tilde{T}$ (solid
line), $\tilde{M}$ (dashed line) and $\tilde{S}$ (dotted line) vs.
$\ln\tilde{\rho_c}$}
\newpage
\DOUBLEFIGURE{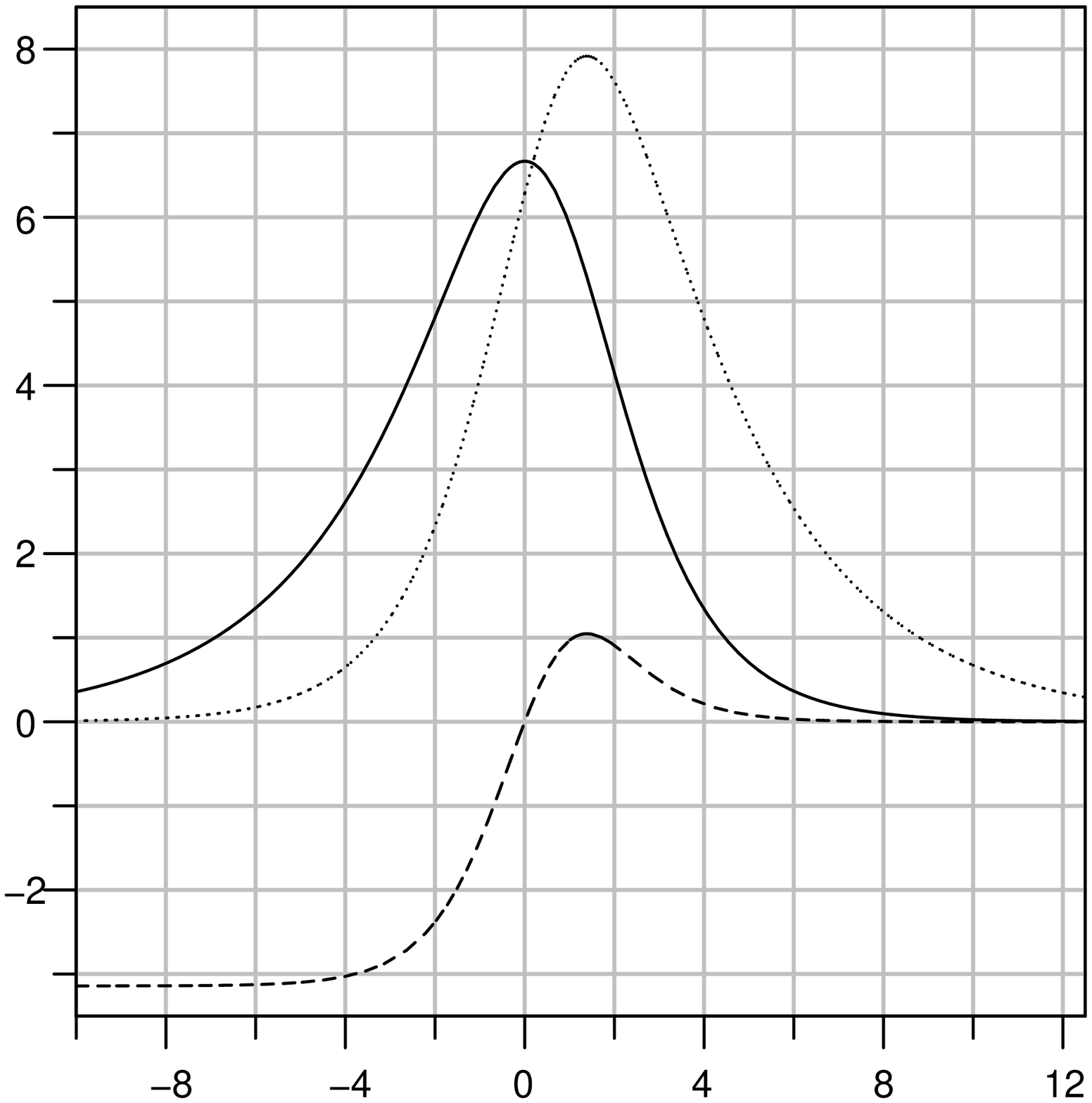,width=2.7in,height=2.7in}{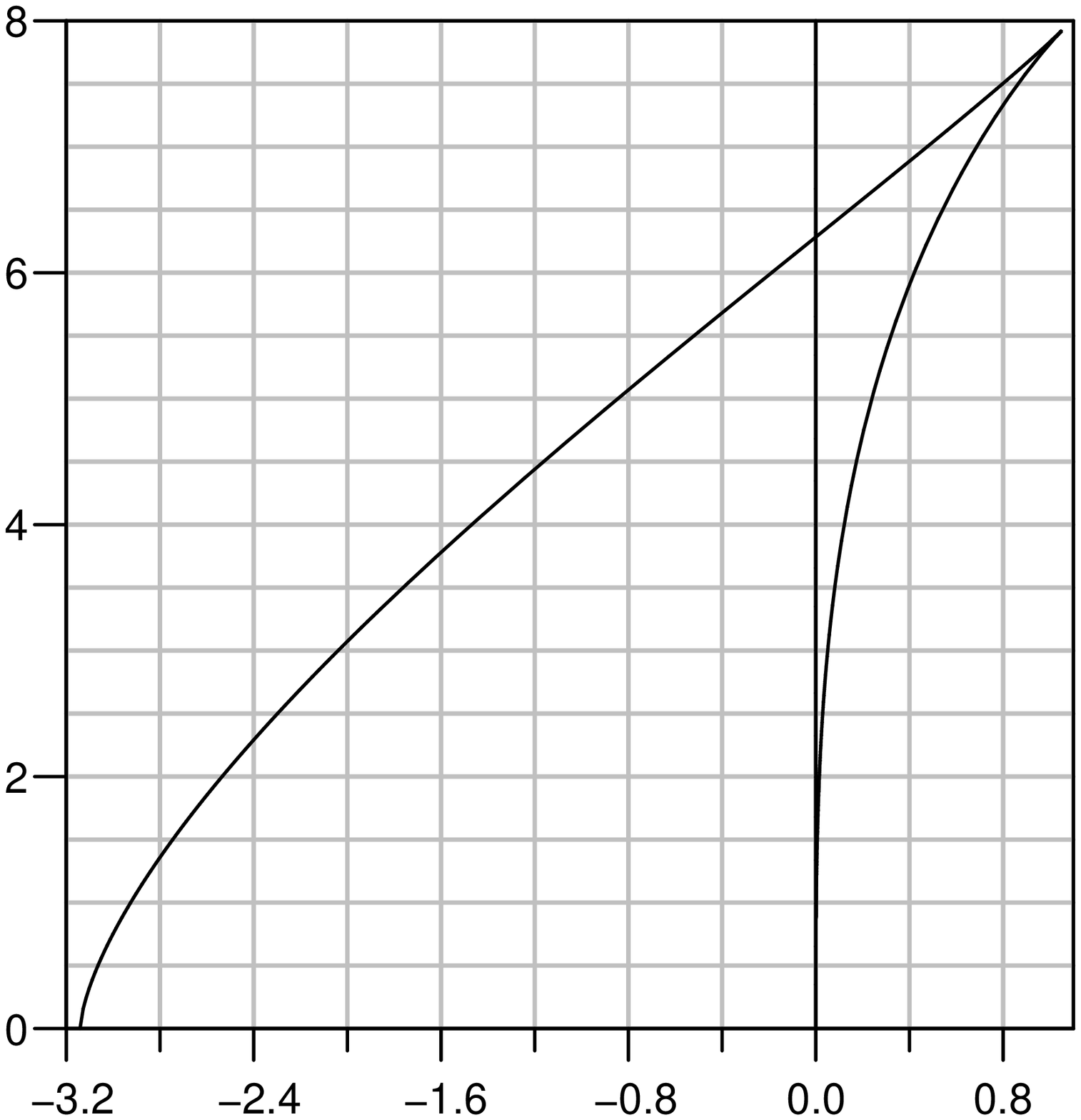,width=3in,height=3in}{\label{fig3}$d=3$:
$10\tilde{T}$ (solid line), $\tilde{M}$ (dashed line) and
$\tilde{S}$ (dotted line) vs.
$\ln\tilde{\rho_c}$}{\label{fig4}$d=3$: $\tilde{S}$ (ordinate) vs
$\tilde{M}$ (abscissa)}

\vspace{3mm}\noindent Our main findings are summarized below:
\vspace{5mm}

\noindent $\bullet\quad \mathbf{4\leq d\leq 10}$. The behaviour is
similar to the $d=4$ case which is illustrated in Fig.~\ref{fig1}.
To the right of their respective peaks, the temperature, mass and
entropy exhibit an infinite series of damped oscillations:
configurations in this range are unstable to one or more radial
modes, although the precise onset of instability depends on the
ensemble \cite{pagephillips}. In the canonical ensemble, at fixed
red-shifted temperature, all configurations to the right of
$T_{max}$ are unstable.  In the microcanonical ensemble, at fixed
mass, all configurations to the right of $M_{max}$ are unstable.
The maximum temperature configuration always occurs at a lower
central density than the maximum mass and entropy configuration
(note that for $d=10$ they lie very close). The numerical values
given in Table~\ref{tab1} for $d=4$ agree with
\cite{pagephillips}.

\vspace{5mm}\noindent $\bullet\quad \mathbf{d\geq 11}$. In this
case the red-shifted temperature, mass and entropy increase
monotonically with the central density, asymptoting to their
maximum values as the central density goes to infinity, see
Fig.~\ref{fig2}. Since the specific heat is always positive, one
would expect these configurations to be stable for all values of
$\rho_c$. To the author's knowledge, this case has not been
studied before.  Note that this critical dimension was found
analytically and not numerically: its exact (non-integer) value is
in fact just under $11$, see Appendix \ref{appendix2}.

\newpage\noindent $\bullet\quad \mathbf{d=3}$. In this special case an
exact solution exists, see Appendix \ref{appendix1}. We found the
following expressions for the (dimensionless) red-shifted
temperature, mass and entropy:
\begin{align}
\tilde{T}&=\frac{\tilde{\rho_c}^{1/3}}{1+\tilde{\rho_c}/2}\\
\tilde{M}&=\frac{\pi(\tilde{\rho_c}-1)}{\left(1+\tilde{\rho_c}/2\right)^2}\\
\tilde{S}&=\frac{3\pi\tilde{\rho_c}^{2/3}}{1+\tilde{\rho_c}/2}\,.
\end{align}
These are plotted in Figs.~\ref{fig3} and \ref{fig4}. Here the
mass is defined such that it vanishes for the massless BTZ black
hole. The minimum mass $M=-1/8G_3$ ($\tilde{M}=-\pi$) corresponds
to $AdS_3$ in global coordinates. The maximum mass is $M=1/24G_3$
($\tilde{M}=\pi/3$). This is the same as the range of masses found
for black holes localized on an $AdS_3$ brane
\cite{ehm2}.\footnote{Fig.~\ref{fig4} should be compared to Fig.~1
in \cite{ehm2} although the dependence of the entropy and
temperature on the other parameters ($G_3$, $l$, $a$) is different
for the two cases. The expressions for the mass, temperature and
entropy in \cite{ehm2} also depend on the parameter $\lambda$,
which is related to the positions of the two branes.} The limiting
infinite central density solution is the same as the massless BTZ
black hole except that there is a curvature singularity
proportional to a $\delta$-distribution supported at the origin
(Appendix \ref{appendix1}). As in the $4\leq d\leq 10$ case, one
would expect that configurations to the right of the peak of
temperature (mass) in Fig.~\ref{fig3} are unstable to radial
perturbations with canonical (microcanonical) boundary conditions.

\section{Comment: validity of perfect fluid approximation}\label{sec4} Under what
conditions is this a good model of quantum fluctuations about the
thermal $AdS$ saddle point (that is, $AdS$ periodically identified
in imaginary time)? For a general (non-conformal,
self-interacting, arbitrary spin) field theory in a static curved
spacetime, the energy-momentum tensor (\ref{eqn:tmunu}) with the
Stefan-Boltzmann relation (\ref{eqn:stefanb}) corresponds to the
leading term in the high-temperature expansion of the one-loop
effective action \cite{dowkerkennedy,kirsten,burgess}. The higher
order terms will be small if the red-shifted temperature satisfies
$T\gg \mathcal{R}^{-1}$ and $T\gg m_{eff}$, where $\mathcal{R}$ is
a characteristic curvature radius of the spacetime and $m_{eff}$
is the effective mass of the field \cite{kirsten}.\footnote{In
general, the effective mass will contain a term involving the
Ricci scalar, as well as terms involving the interaction
potential. However, since we are expanding about the trivial
$\Psi=0$ background, there will not be any interaction terms.} The
constant $a$ is proportional to the effective number of massless
degrees of freedom $g$:
\begin{equation}\label{eqn:freedom}
a=(d-1)\pi^{-d/2}\Gamma\left(\frac{d}{2}\right)\zeta(d)g,
\quad\mbox{where}\quad g=n_B+\left(1-2^{-(d-1)}\right)n_F\,,
\end{equation}
with $n_B$ being the number of boson spin states and $n_F$ the
number of fermion spin states. This may be obtained by integrating
the $d$-dimensional Planck distribution over all frequencies,
taking into account the degeneracy of states factor, see, for
example, \cite{landsbergdevos}. Note that we are also assuming
that the characteristic curvature radius $\mathcal{R}$ is much
greater than the Planck length $l_{Planck}$ so that vacuum
polarization terms and higher order quantum gravity effects are
negligible.

In the context of the configurations studied in this paper, we can
be more precise. If the field is massless (or if the red-shifted
temperature is well above the rest mass), the condition for the
higher order terms in the high-temperature expansion to be small
becomes
\begin{equation}\label{eqn:valid}
T\gg \mbox{max}\{\left(\kappa\rho_c\right)^{1/2},\ l^{-1}\},
\end{equation}
since for $\kappa\rho_c \gtrsim l^{-2}$ the origin will be the
place of maximum curvature.  Now $d\psi/dr>0$ and condition
(\ref{eqn:psinorm}) imply $\psi(0)<0$. From equation
(\ref{eqn:redshifted}) and $V(0)=0$ it follows that
$T_{loc}(0)>T$. Equation (\ref{eqn:valid}) and the
Stefan-Boltzmann relation (\ref{eqn:stefanb}) may then be used to
show that the higher order terms are small if the dimensionless
temperature satisfies
\begin{equation}\label{eqn:boundtemp}
\gamma^{-1/d}\ll \tilde{T}\leq \tilde{T}_{max}\,,
\end{equation}
or if the dimensionless density falls in the range
\begin{equation}\label{eqn:bounddensity}
\gamma^{-1}\ll \tilde{\rho_c}\ll \gamma^{2/d}\,,
\end{equation}
where $\gamma\equiv a^{-1}(l/l_{Planck})^{d-2}$. For $l\gg
l_{Planck}$, as we have assumed, and if the number of degrees of
freedom is not too large, $\gamma$ is a large number and this
model is applicable in a wide domain.\footnote{The upper limits in
(\ref{eqn:boundtemp}) and (\ref{eqn:bounddensity}) are well below
the Planck scale in that case.} In that case, in $3\leq d\leq 10$,
the majority of the configurations corresponding to the curves in
Fig.~\ref{fig1} and Fig.~\ref{fig3} would belong to this regime,
including the maximum temperature configuration. On the other
hand, in $d\geq 11$ (Fig.~\ref{fig2}), the maximum temperature
configuration corresponds to infinite central density and
therefore higher order terms would come into play well before this
point is reached.  It would be interesting to understand better
the significance of this.

The same factor $\gamma$ appears if we compare the maximum
temperature of thermal radiation to the Hawking-Page temperature
$T_{HP}=(d-2)/2\pi l$: we find $T_{max}/T_{HP}\sim \gamma^{1/d}\gg
1$ if $\gamma\gg 1$. This is consistent, since thermal radiation
heated to a temperature $T>T_{max}$ must collapse to a black hole
with a lower free energy.  Similarly, if we follow
\cite{hawkingpage} and consider the microcanonical ensemble at
fixed energy $E$, it may be estimated that an equilibrium of a
black hole with thermal radiation is more probable than thermal
radiation alone if $E>E_1$, where
\begin{equation}
E_1^{2d-3}\sim \kappa^{-d}a^{d-3}l^{(d-1)(d-3)}\,.
\end{equation}
For consistency we require $M_{max}>E_1$ and indeed we find
$M_{max}/E_1\sim \gamma^{(d-3)/(2d-3)}\gg 1$, assuming
$d>3$.\footnote{In $d=3$, there exists a stable equilibrium of a
BTZ black hole and thermal radiation for any $E>0$.} Furthermore,
we may compare the maximum entropy of thermal radiation to the
entropy of a black hole of the same mass,
$S_{BH}=2\pi\Sigma\kappa^{-1}r_{H}^{d-2}$, where $r_H$ is the
horizon radius: we find $S_{max}/S_{BH}\sim\gamma^{-1/d}\ll 1$,
which is consistent with the Bekenstein bound. Note that these
comparisons should not be taken too literally, since we should
really be comparing the radiation values to the black hole values
calculated at the one-loop level.

\section{Conclusions}\label{sec5}
We have studied spherically symmetric equilibria of perfect fluid
radiation with asymptotically $AdS$ boundary conditions in various
dimensions. As in the $d=4$ case, there is a maximum red-shifted
temperature, maximum mass and maximum entropy in all dimensions
$d\geq 3$. We computed these values for $3\leq d\leq 13$. In $d=3$
the analytical results serve as a useful toy model, although it
must be kept in mind that $2+1$ gravity has some special
properties that do not necessarily carry over into higher
dimensions. In $4\leq d\leq 10$ the behaviour is similar to the
$d=4$ case studied in \cite{pagephillips}. In $d\geq 11$ there is
a qualitative difference in behaviour: the temperature, mass and
entropy are monotonic functions of the central density,
approaching their maximum values as the central density tends to
infinity, with no oscillations. Whilst we are not certain of the
significance of the latter result, what we can say is that it
applies in a more general setting than that of a radiation fluid.
The oscillatory profile of thermodynamic variables at high central
densities, associated with modes of instability, is a feature of a
wide class of stellar models: it appears in general relativity
whenever the high density core of a star is described by a
gamma-law equation of state \cite{misnerzapolsky, chavanis2,
relstellstruc}, or indeed certain polytropic equations of state
\cite{tooper}, as well as in Newtonian theory (the isothermal
sphere \cite{chavanis1}), and is independent of the properties of
the outer regions \cite{relstellstruc}. As shown in the dynamical
systems analysis of \cite{relstellstruc}, for an asymptotically
gamma-law equation of state, it can be traced to the nature of the
self-similar solution (\ref{eqn:singular}), which is a fixed point
of the Tolman-Oppenheimer-Volkoff equations. In $d=4$ it is a
stable focus, resulting in oscillatory behaviour in the high
density limit. Their analysis may be extended to $d$ dimensions,
see Appendix \ref{appendix2}. We found that in $4\leq d\leq 10$ it
is a stable focus, whereas in $d\geq 11$ it is a stable node,
resulting in monotonic profiles, as seen in this paper.

One may also look at these solutions from another perspective,
which comes from considering the connection between dynamics and
thermodynamics \cite{sorkinwaldzhang}. The entropy of a given
state is expected to measure the logarithm of the fraction of time
the system spends in that state throughout its dynamical history
\cite{wald}. For matter configurations, this has the implication
that the stable equilibrium configurations should correspond to
local maxima of the entropy at fixed energy
\cite{sorkinwaldzhang}. The relevance of this is as follows.
Firstly, it may be shown that all static spherically symmetric
asymptotically $AdS$ configurations of a radiation fluid, i.e. the
solutions to the Tolman-Oppenheimer-Volkoff equations
(\ref{eqn:tov}) are extrema of the total entropy
(\ref{eqn:entropy}).\footnote{This was shown in
\cite{sorkinwaldzhang} for the $\Lambda=0$, $d=4$ case of
radiation in a box of fixed radius, but it extends to the
asymptotically $AdS$ case.} The converse statement, that the
solutions to equations (\ref{eqn:tov}) are the only extrema of the
entropy, would hold only if it can be assumed that all extrema are
spherically symmetric and that all fluid spacetimes of the type
considered contain a maximal hypersurface \cite{sorkinwaldzhang}.
Existence of maximal hypersurfaces in asymptotically flat
\cite{bartnik} and asymptotically $AdS$ \cite{akutagawa}
spacetimes has been proved subject to certain conditions. The
long-standing conjecture that a static perfect fluid star in $d=4$
is necessarily spherically symmetric appears to have been proved
in \cite{masood} for physically reasonable equations of state in
asymptotically flat space. It is not known if a result of this
kind exists in asymptotically $AdS$ space, in arbitrary dimension.
Note that in the asymptotically $AdS$ case one can consider more
general conformal structures at infinity, the so-called
asymptotically locally $AdS$ spacetimes \cite{skenderis}.  Thus in
this paper we only dealt with a restricted class of solutions.
Secondly, \cite{sorkinwaldzhang} showed that in the $\Lambda=0$,
$d=4$ case of radiation in a box, the criterion for thermodynamic
stability -- that the extremum is a strict local maximum of the
entropy -- is equivalent to the condition for dynamical stability
under linear radial perturbations.  This stability argument would
be expected to generalize to the asymptotically $AdS$ case in $d$
dimensions, although it would be useful to carry out a detailed
perturbation analysis. A similar formulation should also be
possible in the canonical ensemble, where stable equilibria should
correspond to local minima of the free energy at fixed red-shifted
temperature.

Finally, it would be interesting to consider these solutions in
the context of the Karch-Randall model, where one has an
asymptotically $AdS_{d-1}$ brane in an $AdS_{d}$ bulk
\cite{karchrandall,kalopersorbo}. According to $AdS/CFT$,
classical dynamics in the bulk is dual to the quantum dynamics of
a CFT living on the brane, in the planar limit, coupled to
$(d-1)$-dimensional Einstein gravity. The strongly-coupled CFT has
a UV cutoff $\sim 1/L$ and an IR cutoff $\sim 1/l$, where $L$ and
$l$ are respectively the bulk and brane $AdS$ lengths. This theory
has been interpreted as a defect CFT \cite{dewolfe}, although
there are a number of issues still to be resolved. In the original
Karch-Randall model, the bulk volume is infinite and the zero mode
graviton is not normalizable, nevertheless for $l\gg L$ there is
an ultralight graviton with a tiny mass $m^2\sim L^{d-3}/l^{d-1}$
and $(d-1)$-dimensional gravity is effectively reproduced at
distances $r\gtrsim L$ on the brane, although its mass, and the
extra dimension, would become apparent  at very large distances
$r\gtrsim l^{d-2}/L^{d-3}$ \cite{kalopersorbo}.\footnote{For
$L\lesssim r\lesssim l$ the $(d-1)$-dimensional graviton is a
composite made out of the ultralight mode and heavier Kaluza-Klein
modes, whereas for $l\lesssim r\lesssim l^{d-2}/L^{d-3}$ it is
just the ultralight mode \cite{kalopersorbo}.} However, if one
introduces into the bulk a second positive tension $AdS_{d-1}$
brane \cite{kogan,twogravitons}, one recovers the zero mode and
$(d-1)$ dimensional gravity is valid all the way to infinity. One
now has two gravitons: the massless one and the ultralight one,
although the ultralight one decouples in the limit that the second
brane approaches the turn-around point of the warp factor
\cite{twogravitons}. Now if the metric on one of the branes is
given exactly by one of the thermal radiation configurations
studied in this paper, one may wonder what happens in the bulk? In
this context, it is interesting to note that in the $d=4$ version
of this model, with two $AdS_3$ branes in an $AdS_4$ bulk, the
bulk metric is the $AdS$ C-metric studied in \cite{ehm2} and the
other brane contains a localized black hole (Appendix
\ref{appendix1}). It would be interesting to see if this property
is just a special feature of this lower dimensional model or
whether there is a regime where it would hold in higher
dimensions. We note that the issue of the Hawking-Page transition
in this setup has been addressed in \cite{chamblin}, although only
in the limit $l/L\to\infty$ where matter effects can be ignored.

\vspace{8mm} \noindent\textbf{\large{Note}} \vspace{2mm}

\noindent After this work first appeared, two other papers came
out, one by John Hammersley \cite{hammersley} and one by
Pierre-Henri Chavanis \cite{chavanis3}, who have been
independently working on similar problems. Specifically,
\cite{hammersley} studies self-gravitating radiation in
$d$-dimensional $AdS$ and also finds the change in behaviour from
oscillatory in $4\leq d\leq 10$ to monotonic in $d\geq 11$ (note
that the $d=5$ case had previously appeared in \cite{hubeny} in a
study of bulk-cone singularities in $AdS/CFT$). He points out the
possibility that (besides a possible link to string theory!) this
could be related to a similar dimensional transition found in a
cosmological setting \cite{demaret,elskens}, where the analysis of
Belinsky, Khalatnikov and Lifshitz (BKL) \cite{belinsky} was
extended to arbitrary dimension, with the finding that the chaotic
oscillatory dynamics of the gravitational field close to a
spacelike singularity for $4\leq d\leq 10$ is replaced by a
monotonic power law time dependence for $d\geq 11$. It would be
interesting to explore this connection further.

The work of \cite{chavanis3} is a comprehensive study of the
behaviour of general relativistic perfect fluid stars with a
linear equation of state in $d$ dimensional asymptotically flat
space (in a box), including the radiation case, extending the
analysis of \cite{chavanis2} to arbitrary dimension (the Newtonian
case had previously been treated in \cite{sirechavanis}). He also
finds that for $d\geq 11$ the oscillatory behaviour disappears. In
fact the results of Sec. 6.4 in \cite{chavanis3} are consistent
with the findings noted in Appendix \ref{appendix2} of this paper
and provide an alternative derivation of the asymptotic behaviour
of $\Lambda=0$ linear perfect fluid solutions in different
dimensions. The paper \cite{chavanis3} also includes a detailed
stability analysis of the different regimes, with the suggestion
that for these solutions thermodynamic stability is equivalent to
\emph{nonlinear} dynamical stability. The $d=3$ case, treated in
Appendix \ref{appendix1} of this paper, is also analyzed, although
the author suggests that these solutions have finite radial extent
whereas, as noted in Appendix \ref{appendix1}, the finite radial
coordinate at which the density vanishes is at infinite proper
distance from any other point in the spacetime.

\acknowledgments The author would like to thank Stephen Hawking
for many helpful discussions and advice. The author would also
like to thank Claes Uggla for useful correspondence on
self-similarity and James Lucietti and Gian Paolo Procopio for
comments on this manuscript.

\appendix
\section{Exact solution in $d=3$}\label{appendix1}  In this appendix we derive exact solutions
for regular equilibrium configurations of thermal radiation in
$d=3$, both for $\Lambda<0$, and for $\Lambda=0$. Garcia and
Campuzano \cite{garcia} have obtained all static circularly
symmetric perfect fluid solutions in $2+1$ dimensions, and the
solutions below belong to one of the classes presented in that
paper. We found it interesting, however, that the case of a
radiation fluid with $\Lambda<0$ may also be obtained starting
from the $AdS$ C-metric studied by Emparan, Horowitz and Myers
\cite{ehm2}, describing black holes accelerating in $AdS_4$. This
metric, which is a static axially symmetric vacuum solution to the
four-dimensional Einstein equations with negative cosmological
constant, takes the form
\begin{equation}\label{metric:ehm2}
ds^2=\frac{1}{A^2(x-y)^2}\left[H(y)dt^2-\frac{dy^2}{H(y)}+\frac{dx^2}{G(x)}+G(x)d\phi^2\right],
\end{equation}
where
\begin{align}
H(y)&=-\lambda+k y^2-2\mu A y^3 \nonumber\\ G(x)&=1+k x^2-2\mu A
x^3\label{eqn:G}\,,
\end{align}
with $\lambda>0$, $A>0$, $\mu>0$ and $k=-1, \, 0, \,+1$. Points
with $x=y$ correspond to the boundary of the asymptotically
$AdS_4$ geometry, therefore the solution is restricted to the
region $y<x$. The coordinate $\phi$ is an angle and each zero of
$G(x)$ corresponds to an axis for the rotation symmetry, where we
need $G(x)\geq 0$ to preserve the Lorentzian signature. For
$\mu>0$, $G(x)$ has only one positive root $x_2$ and in order to
avoid a conical singularity at this point, the periodicity of the
angular coordinate $\phi$ is set to be
$\Delta\phi=4\pi/|G'(x_2)|$. The smallest zero of $H(y)$ defines
the black hole horizon. See \cite{ehm2} for further details. The
authors of \cite{ehm2} considered putting two positive tension
asymptotically $AdS_3$ branes in this spacetime: one at $x=0$, the
black hole brane, and a second non-singular brane at $y=0$. This
is a lower dimensional version of the two-brane Karch-Randall
model where, at least in a certain range of parameter space, bulk
gravity in $AdS_4$ is expected to reproduce $AdS_3$ gravity on
each brane, plus quantum corrections coming from a CFT. The
different values of $k$ correspond to different slicings of
$AdS_3$.\footnote{In the $\mu=0$ case when the bulk metric
(\ref{metric:ehm2}) is locally $AdS_4$, $k=-1$ corresponds to
global coordinates on the $AdS_3$ slices, whilst $k=0$ and $k=1$
give the metrics of the massless and massive BTZ black holes
respectively \cite{ehm2}.} Now, if there is a quantum corrected
black hole on the first brane, then the other non-singular brane
at $y=0$ should correspond to thermal radiation in $AdS_3$ at
constant red-shifted temperature. This may verified by examining
the induced metric on the surface $y=0$,
\begin{equation}\label{metric:induced}
ds^2=\frac{1}{A^2 x^2}\left[-\lambda
dt^2+\frac{dx^2}{G(x)}+G(x)d\phi^2\right],
\end{equation}
where $x$ is restricted to lie in the range $0<x\leq x_2$. The
Einstein tensor for this $3$-metric is
\begin{equation}\label{eqn:3einstein}
G_{\nu}^{\mu}=A^2\mbox{diag}(1,1,1)+\mu A^3 x^3
\mbox{diag}(-2,1,1),
\end{equation}
which describes a $p=\rho/2$ perfect fluid with density
\begin{equation}\label{eqn:3density}
\rho=\frac{2\mu A^3 x^3}{8\pi G_3}
\end{equation}
in a spacetime with cosmological constant $\Lambda=-1/l^2=-A^2$.
To convert this to the standard BTZ form, that is
\begin{equation}\label{metric:3}
ds^2=-N^2 dt^2 +\left(\frac{r^2}{l^2}-8G_3 m\right)^{-1}dr^2 +r^2
d\theta^2,\quad 0\leq\theta\leq 2\pi\,,
\end{equation}
where $N(r)$ is the lapse function, change variables to the
following:
\begin{align}
r&=\frac{2}{A|G'(x_2)|}\frac{\sqrt{G(x)}}{x}\,,\nonumber\\
\theta&=\frac{|G'(x_2)|}{2}\phi\,.
\end{align}
Observe that $x=x_2$ corresponds to $r=0$, and $x=0$ corresponds
to $r=\infty$. It is not possible to express the functions $m$,
$\rho$ and $N$ in a simple form in terms of $r$, nevertheless, it
is possible obtain explicit expressions for the red-shifted
temperature, mass and entropy.  In terms of the dimensionless
variables (\ref{eqn:dimensions}), we find
\begin{align*}
\tilde{T}&=\frac{\tilde{\rho_c}^{1/3}}{1+\tilde{\rho_c}/2}\\
\tilde{M}&=\frac{\pi(\tilde{\rho_c}-1)}{\left(1+\tilde{\rho_c}/2\right)^2}\\
\tilde{S}&=\frac{3\pi\tilde{\rho_c}^{2/3}}{1+\tilde{\rho_c}/2}\,.
\end{align*}
Note that in the metric (\ref{metric:induced}), the different
values $k=-1$, $k=0$ and $k=1$ are mapped to the ranges
$0\leq\tilde{\rho_c}< 1$, $\tilde{\rho_c}=1$ and
$\tilde{\rho_c}>1$ respectively. Consider now the limit
$\rho_c\to\infty$. Equations (\ref{eqn:G}) and
(\ref{eqn:3density}) require $\mu A\to 0$ and $x_2\approx (2\mu
A)^{-1}\to\infty$. The limiting solution has $\rho=0$ and $m=0$
for all $r> 0$ which is the same as the massless BTZ black hole.
However, there is a curvature singularity proportional to a
$\delta$-distribution supported at the origin.

The $\Lambda=0$ case may be obtained from the metric
(\ref{metric:induced}) by making the transformation $\hat{x}=Ax$,
$\hat{\phi}=A^{-1}\phi$, $\hat{G}=A^2 G$ and taking the limit
$A\to 0$. Dropping the hats, this gives the $3$-metric
\begin{equation}\label{metric:m3}
ds^2=\frac{1}{x^2}\left[-\lambda
dt^2+\frac{dx^2}{G(x)}+G(x)d\phi^2\right],
\end{equation}
where $G(x)=kx^2-2\mu x^3$ and the period of $\phi$ is
$4\pi/|G'(x_2)|$.  The Einstein tensor for this metric is
\begin{equation}\label{eqn:3einstein0}
G_{\nu}^{\mu}=\mu x^3 \mbox{diag}(-2,1,1),
\end{equation}
which describes a $p=\rho/2$ perfect fluid with density
\begin{equation}\label{eqn:3density0}
\rho=\frac{2\mu x^3}{8\pi G_3}\,.
\end{equation}
and $\Lambda=0$. Making the choice $k=1$ (this is the only
consistent choice), we restrict to $0<x\leq x_2=1/2\mu$ so that
$G(x)\geq 0)$. We may compare the metric (\ref{metric:m3}) to a
more standard form for $\Lambda=0$ static solutions in $2+1$
gravity:
\begin{equation}\label{metric:30}
ds^2=-N^2 dt^2+(1-4G_3 m)^{-2}dr^2+r^2 d\theta^2,\quad 0\leq
\theta< 2\pi\,,
\end{equation}
where $N(r)$ is the lapse function.\footnote{This differs from the
usual form \cite{deser, ashtekar} by a trivial change of
variables.} Now in three dimensions  the Weyl tensor vanishes so
the spacetime outside a point mass is locally flat \cite{deser}:
the point mass only affects the spacetime globally, making it into
a cone, the value of the mass being given by the conical deficit
angle at infinity, $\delta_{\infty}=8\pi G_3 m$. Note that this
differs from the mass defined by the metric (\ref{metric:3}) with
$\Lambda=0$. In particular $m=1/4G_3$ in the normalization of
(\ref{metric:30}) corresponds to $m=0$ in the $\Lambda=0$ version
of (\ref{metric:3}). Note also that in the case under
consideration, we are dealing not with point sources but with
smooth perfect fluid configurations. Changing variables to
\begin{align}
r&=\frac{2}{|G'(x_2)|}\frac{\sqrt{G(x)}}{x}=4\mu\sqrt{1-2\mu
x}\,,\nonumber\\
\theta&=\frac{|G'(x_2)|}{2}\phi=\frac{\phi}{4\mu}\,,
\end{align}
we see that $x=x_2$ corresponds to $r=0$, and $x=0$ corresponds to
$r=r_{max}=4\mu>0$. The radial coordinate is bounded: $0\leq
r<r_{max}$. After some algebra we obtain the following expressions
for the mass, density and lapse function of the regular radiation
fluid configurations with $\Lambda=0$:
\begin{align}
m&=\frac{1}{4G_3}\left[1-\left(1-2\pi G_3\rho_c
r^2\right)^2\right]\nonumber\\\rho&=\rho_c\left(1-2\pi G_3\rho_c
r^2\right)^3\nonumber\\ N&=N_0 \left(1-2\pi G_3\rho_c
r^2\right)^{-1}\,,
\end{align}
where $\rho_c=(32\pi G_3\mu^2)^{-1}$ is the central density. The
mass varies from $m=0$ at $r=0$ to $m\to 1/4G_3$ as the radius
$r\to r_{max}$. This is exactly the range of masses one would
expect in $2+1$ gravity with zero cosmological constant with
smooth matter sources \cite{ashtekar}. It may be verified that the
proper distance between two points in the spacetime diverges if
the radial coordinate of one of them approaches $r_{max}$: thus
$r=r_{max}$ lies at infinity and the solutions are unbounded.
There are no conical singularities: the conical deficit angle
increases smoothly from zero at $r=0$ and tends to $2\pi$ as $r\to
r_{max}$. One may visualize the geometry represented by the the
spatial part of the metric (\ref{metric:30}) by embedding it into
a $3$-dimensional flat Euclidean space $(r,\theta,z)$: one ends up
with a semi-infinite cigar, asymptoting to a cylinder at
infinity.\footnote{As noted in \cite{chavanis3}, these unbounded
solutions are marginally stable (as in the $d\geq 4$ case).
However, one may enclose the fluid in a box of finite proper
radius and match it on the outside to a cone to obtain stable
asymptotically conical solutions.} The cylinder is the $d=3$
analogue of the singular self-similar solution
(\ref{eqn:singular}): it is the geometry of a point particle at
the origin of mass $M=1/4G_3$ (in particular, $\rho=0$ for all
$r>0$) \cite{deser}. As in the $d>3$ case, this solution also
corresponds to the $\rho_c\to\infty$ limit of the $\Lambda<0$
radiation fluid configurations for $r/l\to 0$. In that case, the
second of equations (\ref{eqn:tov}) implies that $\rho=0$ for all
$r>0$. This explains our previous observation, that the
$\rho_c\to\infty$ limit of the $\Lambda<0$ radiation fluid
configurations is the same as a massless BTZ black hole for $r>0$.

\section{Phase plane analysis of $\Lambda=0$ scale-invariant
system}\label{appendix2} In this appendix we outline a dynamical
systems analysis of the $d$-dimensional Tolman-Oppenheimer-Volkoff
equations in the $\Lambda=0$ case, which yields an exact
(non-integer) value for the critical dimension at which the
oscillatory behaviour ceases. For generality, we assume a linear
(``gamma-law'') equation of state $p=q\rho$, with $0\leq q\leq 1$:
this leads to the equations being scale invariant, which greatly
facilitates the analysis. We also restrict attention to $d\geq 4$
(the $d=3$ case must be treated separately, we do not describe it
here). In $d=4$, analogous phase portraits have been considered in
\cite{sorkinwaldzhang,collins2,chavanis2} using non-compact
variables. We follow the formulation introduced in
\cite{relstellstruc},  in which the state space of the dynamical
system is compact, i.e. the boundaries are included.

The Tolman-Oppenheimer-Volkoff equations in $d$ dimensions with
$\Lambda=0$ read:
\begin{align}\label{eqn:tovgen}
\frac{dm}{dr}&=\Sigma r^{d-2}\rho\nonumber\\
\frac{dp}{dr}&=-\frac{(d-3)\kappa m\rho}{(d-2)\Sigma
r^{d-2}}\left(1+\frac{p}{\rho}\right)\left(1+\frac{\Sigma
r^{d-1}p}{(d-3)m }\right)\left(1-\frac{2\kappa m}{(d-2) \Sigma
r^{d-3}}\right)^{-1}\,,
\end{align}
where the metric is as before, i.e. (\ref{metric:ansatz}) with
$\Lambda=0$. They have been studied before in \cite{deleon} where
a higher-dimensional version of Buchdahl's limit was derived. We
are interested in the linear and homogeneous equation of state
$p=q\rho$ which is the only equation of state that leaves
(\ref{eqn:tovgen}) invariant under the scale transformation
\begin{equation}\label{scale2}
r \rightarrow C r,\quad \rho\rightarrow C^{-2}\rho,\quad
p\rightarrow C^{-2}p,\quad m\rightarrow C^{d-3}m\,,
\end{equation}
where $C\in\mathbb{R^+}$ \cite{collins2, relstellstruc}. In that
case, introducing the dimensionless variables\footnote{These may
be regarded as a relativistic analogue of the Milne variables,
homology invariant variables in the Newtonian theory. If the speed
of light $c$ is put back in, we have for a linear equation of
state $p=q\rho c^2$, where $\rho$ is now the rest frame mass
density. The Newtonian limit is given by $c\to\infty$, $q\to 0$
with $q c^2$ constant. The Einstein equations then reduce to an
Emden equation describing an isothermal sphere in Newtonian
gravity \cite{chavanis2}. Note that in the Newtonian case a
polytropic equation of state also preserves scale invariance
\cite{relstellstruc}.}
\begin{equation}\label{eqn:uvdefine}
u=\frac{\Sigma r^{d-1}\rho}{m},\quad v=\frac{\rho }{p}\frac{\kappa
m}{(d-2)\Sigma r^{d-3}}\left(1-\frac{2\kappa m}{(d-2)\Sigma
r^{d-3}}\right)^{-1}\,,
\end{equation}
equations (\ref{eqn:tovgen}) are transformed (cf.
\cite{relstellstruc}) to the following autonomous system:
\begin{align}\label{eqn:auton}
\frac{du}{d\xi}&=-u \left( 1-d+u+ \left( d-3+qu \right)  \left(
1+q \right) v \right)\nonumber\\ \frac{dv}{d\xi}&=-v \left(
d-3-u+2\,q \left( d-3-u \right) v \right)\,,
\end{align}
where $\xi=\ln r$.  Note that if $m$ and $\rho$ are positive and
the spacetime is static, $u$ and $v$ lie in the ranges $u>0$,
$v>0$. In that case, following \cite{relstellstruc}, we change
from $(u,v)$ to the bounded variables $(U,V)\in (0,1)^2$
\begin{equation}\label{eqn:UVdefine}
U=\frac{u}{1+u},\quad V=\frac{v}{1+v}\,,
\end{equation}
and introduce a new independent variable $\lambda$,
\begin{equation}\label{eqn:deflambda}
\frac{d\lambda}{d\xi}=\frac{1}{(1-U)(1-V)}\,,
\end{equation}
whence equations~(\ref{eqn:auton}) yield the following dynamical
system:
\begin{align}\label{eqn:system}
\frac{dU}{d\lambda}&=U \left( 1-U \right)  \left( d-1-d U- \left(
2\,d-4+ \left( d-3 \right) q \right) V+ \left( 2\,d-3+ \left(
d-q-4 \right) q \right) U V \right)\nonumber\\
\frac{dV}{d\lambda}&=V(1-V)\left(1+(2q-1)V\right)\left(3-d+(d-2)U\right).
\end{align}
As discussed in \cite{relstellstruc}, the asymptotic properties of
a dynamical system are determined by the boundaries, therefore it
is essential to consider the compactified state space $[0,1]^2$.
This requires the system to be $\mathcal{C}^1$ on $[0,1]^2$, which
is clearly the case for the polynomial system (\ref{eqn:system}).

The idea is to solve the dynamical system (\ref{eqn:system}) for
$U(\lambda)$, $V(\lambda)$ on the state space $[0,1]^2$ and then
to obtain the physical perfect fluid configurations $m(r)$,
$\rho(r)$ one inverts the transformations (\ref{eqn:uvdefine}),
(\ref{eqn:UVdefine}), see \cite{relstellstruc}.

A pivotal role in any dynamical systems analysis is played by the
fixed points.\footnote{The Hartman-Grobman theorem states that the
eigenvalues and eigenvectors of the linearization of a dynamical
system about a hyperbolic fixed point completely characterize the
flow of the system in the neighbourhood of this point, that is,
the flow of the full non-linear system is topologically equivalent
to the flow of its linearization in this neighbourhood.} The fixed
points of (\ref{eqn:system}) in the unit square $[0,1]^2$ and
their associated eigenvalues are listed in Table~\ref{tab2}, where
the parameters $\omega_1$ and $\omega_2$ are given by
\begin{align}\label{eqn:tolmaneigen}
\omega_1&={\frac { \left( d-3 \right)  \left( 1+q \right) \left(
d-3+
 \left( d-1 \right) q \right) }{ 2\left( d-2 \right)  \left( d-1+
 \left( d-3 \right) q \left( 2+q \right)  \right) }}\nonumber\\
 \omega_2&={\frac { \left( d-3 \right)  \left( 1+q \right) \sqrt {(1+q)^2d^2-2(1+q)(7+5q)d+25q^2+22q+33}}{2 \left( d-2 \right)  \left( d-1+ \left( d-3 \right) q \left(
2+q \right)  \right) }}.
\end{align}

\TABULAR{@{\extracolsep{\fill}}|c|cc|c|}{\hline & & & \\ Fixed
point & $U$ & $V$ & Eigenvalues\\ \hline & & & \\ $T_1$ & $0$ &
$0$ & $(d-1)\ ,\ -(d-3)$\\ $T_2$ & $\frac{d-1}{d}$ & $0$ &
$-\frac{(d-1)}{d}\ ,\ \frac{2}{d}$\\ $T_3$ & $\frac{d-3}{d-2}$ &
$\frac{2}{d-1+(d-3)q(2+q)}$ & $-\omega_1\pm\omega_2$\\$T_4$ & $1$
& $0$ & $1\ ,\ 1$\\ $T_5$ & $1$ & $1$ & $q(1+q)\ ,\ -2q$\\ $T_6$ &
$0$ & $1$ & $-(d-3)(1+q)\ ,\ 2(d-3)q$\\ \hline}{\label{tab2}Fixed
points of (\ref{eqn:system}) in the unit square $[0,1]^2$ and
associated eigenvalues, assuming $0\leq q\leq 1$ and $d\geq 4$.}

The global dynamics of the system is determined by its past and
future limit sets.  For the system (\ref{eqn:system}) it takes a
particularly simple form: it may be shown (cf.
\cite{relstellstruc}) that orbits in the interior of the unit cube
originate either from the fixed point $T_2$ or from the fixed
point $T_4$, and converge to the fixed point $T_3$.\footnote{This
follows from local analysis and the monotonicity principle,
whereby the existence of a monotone function on the state space
precludes periodic or recurrent orbits, see \cite{relstellstruc}.
It may be verified that a monotone function is given by $Z=(q
h_1+k_1)^2 (u v)^{-p_1}(1+2q v)^{p_1-1}$ where $h_1=(1+q)(d-3+q
u)v$, $k_1=d-3+(d-1)q$ and
$p_1=4q^2/\left((1+q)(d-3+(d+1)q\right)$, cf. \cite{relstellstruc}
for $d=4$.} The central role is played by the fixed point $T_2$,
corresponding to Minkowski space in spherically symmetric form,
and by the fixed point $T_3$, which represents the singular
self-similar solution (cf. eqn.~(\ref{eqn:singular}))
\begin{equation}\label{eqn:singulargen}
\rho=\alpha r^{-2},\quad m=\alpha\Sigma r^{d-3}/(d-3)\quad
\mbox{where}\quad
\alpha=\frac{2(d-2)(d-3)\kappa^{-1}}{d(d-4)(1+q)^2+3q^2+6q+7}.
\end{equation}

Scale invariance implies that the entire set of solutions with a
regular centre is represented \cite{relstellstruc} in the state
space by a single orbit originating from $T_2$ (as
$\lambda\to-\infty$ or $r\to 0$) and converging to $T_3$ as (as
$\lambda\to\infty$ or $r\to\infty$).\footnote{The one-parameter
set of orbits originating from the fixed point $T_4$ may be
continued to the negative mass domain $u<0$, $v<0$ and describes
solutions with a negative mass singularity surrounded by positive
energy density that enter the state space when they have gained
enough mass \cite{relstellstruc}.} The exact form of this orbit
must be found numerically, however, one may write down asymptotic
approximations for it by linearizing the dynamical system about
the points $T_2$ and $T_3$: these are determined by the
corresponding eigenvalues (and associated eigenvectors) in
Table~\ref{tab2}. It follows from (\ref{eqn:singulargen}) that the
regular perfect fluid configurations with $\Lambda=0$ have
infinite mass and are thus not asymptotically flat: to obtain
finite mass perfect fluid solutions one must confine the radiation
to an unphysical box. Scale invariance then implies that the
regular orbit admits a dual interpretation \cite{sorkinwaldzhang},
whereby different initial segments (equivalently end points) of
the orbit are interpreted as representing distinct regular
equilibria of a perfect fluid in a box of fixed radius $R$, that
may be parametrized by their central density $\rho_c$ (a monotonic
function of $\lambda$), so that $T_2$ represents the solution with
$\rho_c=0$, i.e. an empty box, and $T_3$ is the limiting solution
as $\rho_c\to\infty$.

The crucial fact is that the behaviour of the regular orbit as
$\lambda\to\infty$, i.e. in the neighbourhood of the fixed point
$T_3$, reflects itself in the high-density behaviour of various
asymptotic quantities, such as the mass, the red-shifted
temperature or the entropy, of a linear (i.e. $p=q\rho$) perfect
fluid in a box. Indeed, it was shown in \cite{relstellstruc} that
the equation of state may be quite different away from the centre,
as long as it is still asymptotically linear at high densities.
They considered a fairly general class of equations of state that
are asymptotically linear at high densities and asymptotically
polytropic at low densities and proved that the mass-radius
diagram of these stellar models has a spiral structure (for low
enough polytropic index the radius is finite without the need for
a box).\footnote{The presence of another, non-scale-invariant,
variable in the equations means one has to work with a
$3$-dimensional dynamical system, i.e. the state space is enlarged
to a cube \cite{relstellstruc}. The key is again scale invariance,
which is now only present asymptotically \cite{relstellstruc}.}
The case treated in this paper, where the linear equation of state
was modified by adding a negative cosmological constant, giving
finite mass unbounded configurations without the need for a box,
may be treated similarly.\footnote{For the case of positive
cosmological constant, see \cite{nilssonuggla}.}

The behaviour of the regular orbit near $T_3$ is determined by the
eigenvalues $-\omega_1\pm\omega_2$, where $\omega_1$ and
$\omega_2$ are given by equation (\ref{eqn:tolmaneigen}). These
depend on the value of $q$ and on the dimension $d$. For $0\leq
q\leq 1$ and $4\leq d<10$, $\omega_2$ is always imaginary, and the
fixed point $T_3$ is a stable focus (since $\omega_1$ is
positive).  This leads to oscillatory behaviour, e.g. the spiral
structure theorem in \cite{relstellstruc} or the $4\leq d\leq 10$
case in this paper. However, for a given $0\leq q\leq 1$, there is
a critical dimension $d_{crit}$ above which $\omega_2$ is real and
strictly less than $\omega_1$, i.e. $T_3$ is a stable node and
monotonic behaviour emerges instead. It may be verified that the
critical dimension is always in the range $10\leq d\leq 11$.  The
case of perfect fluid radiation ($q=1/(d-1)$) treated in this
paper gives a critical dimension $d_{crit}=10.96404372..$.
Curiously, a stiff fluid ($q=1$) gives $d_{crit}=10$, and
pressureless dust ($q=0$) gives $d_{crit}=11$. The last result
also corresponds to the Newtonian limit (the isothermal sphere)
and has been obtained before in \cite{sirechavanis}.

\end{document}